\documentclass[prd,amsmath,twocolumn]{revtex4-2}
\usepackage{graphicx}
\usepackage{float}
\usepackage{epstopdf,cancel}
\usepackage{epsf,latexsym,bbm,euscript}
\usepackage{amssymb,amsmath}
\usepackage{mathtools} 
\usepackage{times,graphics}
\usepackage{soul,xcolor}
\usepackage{mathtools}

\def\6{{\langle}}
\def\9{{\rangle}}

\newcommand{\be}{\begin{equation}}
\newcommand{\ee}{\end{equation}}
\newcommand{\ba}{\begin{eqnarray}}
\newcommand{\ea}{\end{eqnarray}}

\def\etal{\textit{et al.}}

\usepackage{url,hyperref}
\hypersetup{colorlinks,linkcolor={blue!55!black},citecolor={red!50!black},urlcolor={blue!45!black}}

\begin{document}

\title{Surface gravity from tidal acceleration}

\author{Pravin Kumar Dahal}
\email{pravin-kumar.dahal@hdr.mq.edu.au}
\affiliation{School of Mathematical \& Physical Sciences, Macquarie University}

\date{\today}

\begin{abstract}

Surface gravity plays a pivotal role in the characterization of black holes and also in formulating the laws of black hole thermodynamics. Despite its significance, defining surface gravity in general spacetimes is a nontrivial task and thus has multiple definitions that lack equivalence in dynamical scenarios. This paper reviews different notions of dynamical surface gravity and then proposes a new definition based on tidal acceleration, which is an alternative way of characterizing spacetime curvature. By integrating tidal acceleration from the horizon to infinity, we could retrieve surface gravity in familiar situations of Schwarzschild and Kerr spacetimes. We outline a generic procedure for calculating surface gravity and substantiate our proposal by investigating the surface gravity of stationary spacetimes and reproducing the results from the literature. Furthermore, we examine surface gravity in nonstationary spacetimes, with a focus on Vaidya spacetime as a model of evaporating black holes.
    
\end{abstract}

\maketitle

\section{Introduction}

Surface gravity is a crucial concept in black hole physics, and its importance is evident in various aspects, including the laws of black hole thermodynamics~\cite{w1,ep16,hw3}. It roughly gives a force required to hold the particle of unit mass at the horizon by an observer at infinity (by means of an infinitely long and massless string)~\cite{ep16,fn17}. However, defining surface gravity, in general, is not straightforward and several definitions of surface gravity exist in the literature, which are not equivalent in dynamical spacetimes~\cite{csv3}. One of the commonly used definitions of surface gravity is the Killing horizon surface gravity, which measures the inaffinity on the Killing horizon, that is, the failure of the geodesic Killing vector to be affinely parametrized. However, this definition is not applicable to spacetimes where the notion of the Killing horizon fails to exist. The Kodama vector~\cite{kd5} provides an alternative approach to defining surface gravity, which is always available in spherical symmetry. The Kodama-Hayward definition of surface gravity, proposed by Hayward~\cite{hw4}, agrees with surface gravity on the horizon of a Reissner-Nordstrom black hole but not with other dynamical constructs. This definition is commonly used in the Hamilton-Jacobi approach to studying Hawking radiation for time-dependent horizons~\cite{hw5,vac7}. For a comprehensive list of the definitions of surface gravity, including their domains of applicability, advantages, and limitations, please see the references above. We will discuss some of these definitions below and provide additional references. We want to mention that owing to limitations in its existing definitions, the appropriate definition of surface gravity in non-trivial backgrounds is still an open question.

Geodesic deviation refers to the way in which the paths of two nearby particles can diverge or converge due to the curvature of space-time. This deviation in geodesic trajectories of test particles can be attributed to the tidal force, resulting in tidal acceleration between them~\cite{MTW:73,fn17}. Thus, in general relativity, tidal acceleration experienced by two neighbouring test particles when they are separated by a small distance is a manifestation of the curvature of spacetime or gravity. Therefore it is interesting to explore how surface gravity can be obtained from the expression for tidal acceleration.

Let us consider timelike geodesics whose tangent at the points on the Cauchy surface is a timelike vector field $l^\alpha$. Then the inward radial component of the rate of change of the vector joining any two nearby geodesics that are radially displaced from one another on the Cauchy surface is given by $R_{\alpha\beta\mu\nu} l^\alpha n^\beta l^\mu n^\nu$, where $R_{\alpha\beta\mu\nu}$ is the Riemann tensor, and $n^\alpha$ is the unit normal to the Cauchy surface. As one intuitively expects, this geodesic deviation is governed by the effective total mass of the isolated system, and one might also expect $R_{\alpha\beta\mu\nu} n^\beta n^\nu$ to contain information about this mass~\cite{aa5}. This makes the objective of extracting surface gravity from tidal acceleration reasonable, and we perform this task here.

The Hawking effect~\cite{hw8} is a well-known phenomenon in the study of black holes, which implies that a distant inertial observer would see the collapsing source radiate. The e-folding relation between the Eddington-Finkelstein coordinate $u$ and the Kruskal coordinate $U$ is necessary to describe the Hawking effect, which suggests that the spacetime curvature and the likely position of the formation of the horizon are essential features for Hawking radiation~\cite{blsm6}. The essential features, namely spacetime curvature is incorporated in the quantity tidal acceleration, which can be integrated from the horizon to infinity, showing that the Hawking effect has its origin at the horizon and is perceived by the observer at infinity. Thus, expressing Hawking's temperature in terms of tidal acceleration is desirable as it reveals particle production as a feature of curved spacetime. Hawking's temperature, in some familiar cases, is proportional to the surface gravity~\cite{hw8,vac7}. In that sense, it is natural to expect the relation for surface gravity in terms of the tidal force.

The effects of particle production in the gravitational field have been closely examined and compared with particle production in an electromagnetic field~\cite{fn17,bb10}. Although there is still no universal consensus on the radiation from the uniformly accelerated charge, the notion of the relative acceleration between the observer and the charge causing the emission of the radiation~\cite{rl9,bw10,hs1} is widely accepted. The relative acceleration causes curvature on the electric field lines generating a stress and the energy required to overcome this stress is emitted as the radiation. Accordingly, one can demand an analogous mechanism for the emission of radiation in gravitational field, which, although is observer dependent, fundamentally depends on the spacetime curvature itself.

This article is organized as follows. We will review dynamical surface gravity in Section~\ref{dsg}. Then in Section~\ref{s2}, we will discuss about the calculation of the surface gravity from tidal acceleration. Specifically, section~\ref{s2a} presents a couple of specific examples to demonstrate that surface gravity can be recovered from tidal acceleration. Subsequently, in Section~\ref{fd}, we provide a formal definition of surface gravity and establish its connection to the usual acceleration due to gravity in the Newtonian limit. Using our proposal, we outline the generic procedure for calculating surface gravity in Section~\ref{s3}. Additionally, we present an illustration using the general spherically symmetric metric in Schwarzschild coordinates. We delve into the realm of stationary spacetimes in Section~\ref{s4}, for which surface gravity is unambiguously defined. Reproducing well-established results, in this case, serves as a pivotal benchmark for assessing the validity and effectiveness of our proposal. In Section~\ref{s5}, we extend the application of our definition to the domain of nonstationary spacetimes. We specifically focus on Vaidya spacetime, which serves as an excellent model of evaporating black holes~\cite{pf1}. Finally, in Section~\ref{s6}, we discuss our results and conclude this article.

We consider the metric $g_{\mu\nu}$ of signature $\left(-,+,+,+\right)$ and take $G=c=1$. Similarly, depending on the convenience, we denote the covariant derivative by $\nabla_\mu$ or by  $D/Dx^\mu$. $\gamma$ denotes the timelike geodesic curve parameterized by $\tau$, $(l^\mu, n^\mu, u^\mu, v^\mu)$ denotes the parallel propagated tetrad along this curve (except in Section~\ref{dsg}) and $e^\mu_{~i}=(n^\mu, u^\mu, v^\mu)$. Greek letters denote tensor indices, and Roman letters denote tetrad indices; in particular, $\hat \tau$ and $\hat \rho$ denote the tetrad indices associated with the components $l^\mu$ and $n^\mu$, respectively. The sign convention used here is adopted by \cite{MTW:73}.

\section{Dynamical surface gravity} \label{dsg}

An observer at infinity from the background of a stationary black hole observes the thermal radiation (also known as Hawking radiation) of temperature $\kappa/2\pi$, $\kappa$ being the black hole's surface gravity~\cite{hw8,w1,fn17}. However, unlike in stationary spacetimes, there is no unambiguous definition of surface gravity in dynamical spacetimes~\cite{fr41,v48}. This is because there is no Killing vector field that is timelike at infinity in such spacetimes. This vector field and the associated Killing horizon is the basis for the definition of the (Killing horizon) surface gravity in stationary spacetimes. Various proposals have been made for generalizing surface gravity to dynamical cases, which reduce to the existing definition in stationary spacetimes. Even for sufficiently slowly evolving horizons with properties sufficiently close to their classical counterparts, these generalizations are practically indistinguishable~\cite{ny40}.

\subsection{Inaffinity surface gravity}

As horizon is the basis for the definition of surface gravity, some notion of quasi-local horizon is necessary for its generalization in dynamical spacetimes. One could use the familiar notion of marginally trapped surface which is a compact spacelike two-surface with vanishing expansion of future directed outgoing (or ingoing) null geodesics~\cite{ak52,v48}. For outgoing/ingoing null geodesics $l^\mu$ (or $n^\mu$), marginally outer trapped surface is given by the condition $\theta_l=0$, where
\begin{equation}
    \theta_l= h^{\mu\nu} \nabla_\mu l_\nu, \quad h_{\mu\nu}= g_{\mu\nu}+ l_\mu n_\nu+ l_\nu n_\mu. \label{mts1}
\end{equation}
Then, analogous to the Killing horizon, we can define surface gravity for the non-Killing horizon from the fact that $l^\mu$ is typically non-affinely parametrized, even though it is not necessarily the horizon generator. As the expansion of outgoing null curves on the marginally outer trapped surface vanishes, these curves are locally constant radius orbits as they cross it
\begin{equation}
    l^\nu \nabla_\nu l^\mu= \kappa l^\mu.
\end{equation}
This definition of surface gravity depends only on the location of the quasi-local horizon and is thus quasi-local. However, its value changes on changing the parametrization of the null vector $l^\mu$
\begin{equation}
    l^\mu \to l'^\mu= \frac{dx^\mu}{d\lambda'}= l^\mu \frac{d\lambda}{d\lambda'}.
\end{equation}
This reparametrization changes the surface gravity as
\begin{equation}
    \kappa \to \kappa'= \frac{d\lambda}{d\lambda'} \kappa+ l^\nu \nabla_\nu \left(\frac{d\lambda}{d\lambda'} \right).
\end{equation}
There always exists a choice of parameter such that $l^\mu$ is affinely parametrized and $\kappa=0$. There are several ways to fix the parameter to obtain the unique surface gravity~\cite{ny40,fr41,bf42}, and we will discuss a couple of them here. Note that the parametrization of the Killing vector is fixed by the proper time of a stationary observer at infinity.

\subsubsection{Fodor \etal ~proposal}

Fodor \etal~\cite{fd37} proposes the normalization condition $n^\mu t_\mu= -1$ for ingoing null geodesics, where $t^\mu$ is the asymptotic time translation Killing vector. So, this definition is restricted to the spherically symmetric asymptotically flat spacetimes. Outgoing null geodesics $l^\mu$ is uniquely determined in a non-local manner by this normalization procedure and the relation
\begin{equation}
    l^\nu \nabla_\nu l_\mu= \kappa l_\mu, \label{insg5}
\end{equation}
can be used to define the surface gravity. As $t^\mu$ can be expanded in terms of $l^\mu$ and $n^\mu$ in spherical symmetry, the normalization condition is equivalent to $l^\mu n_\mu=A(t,r)$, where $A(t,r)$ is an arbitrary function. Readers are also referred to the Ref.~\cite{nv38} for the explicit calculations of inaffinity surface gravity. Note that, differing normalization by a constant factor does not affect the value of surface gravity.

\subsubsection{Aberu-Visser proposal}


Let us illustrate the normalization condition proposed by Aberu and Visser~\cite{av15} using general spherically symmetric metric
\begin{equation}
    ds^2= - e^{2 h(t,r)} f(t,r) dt^2+ \frac{1}{f(t,r)} dr^2+ r^2 d\Omega^2. \label{sc6}
\end{equation}
Corresponding null vectors satisfying the normalization condition $l^\mu n_\mu= f/2$, which is proportional to the squared norm of the Kodama vector $K^\mu= e^{-h} (1,0,0,0)$, are
\begin{align}
    l^\mu=& \frac{1}{2} (e^{-h}, f, 0, 0),\\
    n^\mu=& -\frac{1}{2} \left(e^{-h}, - f, 0, 0\right).
\end{align}
Using Eq.~\eqref{insg5}, one could calculate the inaffinity surface gravity
\begin{equation}
    \kappa_1= \frac{1}{2}\left(\frac{\partial f}{\partial r}+ f \frac{\partial h}{\partial r}\right). \label{k19}
\end{equation}
Ref.~\cite{v39} argues that this surface gravity is not the correct even in the static limit, as it misses a factor of $e^h$. This argument is based on the calculation of Hawking temperature in static situations using the standard Wick rotation, demanding the absence of any conical singularity and interpreting in terms of a periodicity in imaginary time. Again, null vectors satisfying the normalization condition $l^\mu n_\mu= e^{2 h} f/2$, which is proportional to the squared norm of the asymptotic time translation Killing vector $t^\mu= (1,0,0,0)$, are
\begin{align}
    l^\mu=& \frac{1}{2} (1, e^{h} f, 0, 0),\\
    n^\mu=& -\frac{1}{2} \left(1, - e^{h} f, 0, 0\right).
\end{align}
As above, one could calculate the inaffinity surface gravity from Eq.~\eqref{insg5}
\begin{equation}
    \kappa_{2l}= \frac{1}{2}\left(e^h \frac{\partial f}{\partial r}+ 2 e^h f \frac{\partial h}{\partial r}+ \frac{\partial h}{\partial t}\right).
\end{equation}
We could also calculate the inaffinity surface gravity corresponding to the null vector $n^\mu$ by using the same formula~\eqref{insg5}
\begin{equation}
    \kappa_{2n}= \frac{1}{2}\left(e^h \frac{\partial f}{\partial r}+ 2 e^h f \frac{\partial h}{\partial r}- \frac{\partial h}{\partial t}\right).
\end{equation}
The Aberu-Visser proposal is to average the surface gravity due to both future and past pointing null vectors
\begin{equation}
    \kappa_2= \frac{\kappa_{2l}+ \kappa_{2n}}{2}= \frac{1}{2}\left(e^h \frac{\partial f}{\partial r}+ 2 e^h f \frac{\partial h}{\partial r}\right).
\end{equation}
This quantity has the correct factor of $e^h$ and thus gives the desired static limit (at the horizon).

\subsection{Hayward-Kodama surface gravity}

In dynamical spherically symmetric spacetimes, Kodama vector provides a generalization to the Killing vector and is parallel to the Killing vector, whenever it exists. Let $r$ be the areal radius of a 4-dimensional general spherically symmetric spacetime (for example, in Eq.~\eqref{sc6}) and $\epsilon_{ab}$, \{$a,b\}= \{0,1$\} be the volume element of the surface orthogonal to the 2-sphere of symmetry. Then the Kodama vector could be defined as~\cite{kd5}
\begin{equation}
    K^a= \epsilon^{ab} \nabla_b r, \quad K_\theta=0=K_\phi. \label{kod15}
\end{equation}
As the Kodama vector is divergence free $\nabla_a K^a=0$, there exists a conserved current $J_a= G_{a b} K^b$. The charge associated with this current is the Misner-Sharp mass~\cite{hw45}.

Using the Kodama vector, Hayward defined the surface gravity on the marginally trapped surface (introduced in Eq.~\eqref{mts1}) as~\cite{kd5, Hayward:2008}
\begin{equation}
    K^a \left(\nabla_b K_a- \nabla_a K_b\right)= \kappa K_b.
\end{equation}
Because of the uniqueness of the Kodama vector, Hayward-Kodama surface gravity is unique. Using Eq.~\eqref{kod15}, we can write this expression for the surface gravity differently as~\cite{hw4}
\begin{equation}
    \kappa= \frac{1}{2} ~\mathrm{div}~(\mathrm{grad}~ r),
\end{equation}
where the divergence and gradient is on the 2-dimensional space orthogonal to the 2-sphere of symmetry. For the general spherically symmetric metric of Eq.~\eqref{sc6}, we have the Kodama vector $K^\mu= e^{-h} (1,0,0,0)$. So, the Hayward-Kodama surface garvity is
\begin{equation}
    \kappa= \frac{1}{2} \frac{\partial f}{\partial r}.
\end{equation}
As mentioned below Eq.~\eqref{k19}, this quantity is missing the factor of $e^h$ to be correct even in the static limit.

There is also the notion of trapping gravity proposed by Hayward~\cite{hw3}. Trapping gravity does not give the correct result for Reissner-Nordström black holes~\cite{ny40}. As a natural generalization of the trapping gravity in spherical symmetry, Mukhoyama and Hayward also proposed another definition of surface gravity~\cite{mh44}. We will not discuss these in detail here (see Refs.~\cite{ny40, fr41,hw3,mh44, pr45}).

\subsection{Peeling surface gravity}

Peeling off properties is the approximate exponential relation (near the horizon) between the affine parameters on the null generators of past and future null infinity. Refs.~\cite{csv3,blsm6,blsm47} argue that Hawking temperature is related to this property of null geodesics that actually reach future null infinity. Associated with the peeling off properties is peeling surface gravity.

In general spherically symmetric spacetimes, radial null geodesics satisfy
\begin{equation}
    \frac{dr}{dt}\approx \pm 2 \kappa \left(r- r_H(t)\right)+  {\cal O} \left(r- r_H(t)\right)^2,
\end{equation}
where $\kappa$ is the peeling surface gravity. More explicitly, for the metric of Eq.~\eqref{sc6}, radial null geodesics take the form
\begin{align}
    \frac{dr}{dt}= & \pm e^{h} f \nonumber \\
    = & \pm \left( e^{h} \frac{\partial f}{\partial r}+ e^{h} f \frac{\partial h}{\partial r}\right) \left(r- r_H(t)\right)+  {\cal O} \left(r- r_H(t)\right)^2.
\end{align}
So, for this spacetime
\begin{equation}
    \kappa= \frac{1}{2} \left( e^{h} \frac{\partial f}{\partial r}+ e^{h} f \frac{\partial h}{\partial r}\right).
\end{equation}
Hence, the peeling surface gravity contains the correct factor of $e^h$ to give the desired static limit.

Surface gravity of dynamical spacetimes that are conformal transforms of stationary spacetimes can be calculated by the procedure outlined in Ref.~\cite{tj3} (see also, Ref.~\cite{mr53,jf54}). It states that surface gravity is invariant under conformal transformations of stationary spacetimes (with nonvanishing conformal factors). There are works on the calculation of various notions of surface gravity in momentum dependent Schwarzschild black hole~\cite{rl49} and Einstein-aether black holes~\cite{dl56}, which are classes of Lorentz invariance violating theories~\cite{m55}. Similarly, Ref.~\cite{bms51} has used the peeling surface gravity to calculate the Hawking temperature of noncommutative Schwarzschild black hole using the tunnelling procedure. Also, calculations of the surface gravity of black holes in Horndeski theory have been performed in Ref.~\cite{hlsv50} and on Lovelock gravity in Ref.~\cite{ms58}. The comparison of different surface gravity definitions in different regimes may provide a clue to the most useful notion of surface gravity in general spacetimes.

Various authors have also proposed new procedures for calculating surface gravity, which have so far been tested only for stationary spacetimes. For example, Ref.~\cite{bl59} presents a definition of surface gravity in terms of the invariant four-volume of a black hole (which is the volume of a spacetime interior to the horizon) that has formed from a gravitational collapse in a finite proper time. Similarly, Ref.~\cite{jp60} defines surface gravity in terms of the two-dimensional expansion of rate of the geodesic congruence as seen by freely falling observers crossing the horizon. Additionally, Ref.~\cite{cekl61} proposes a method for calculating the surface gravity from the entropic force on holographic screens~\cite{v62} for spherically symmetric black holes.

\section{Tidal surface gravity} \label{s2}

\subsection{Schwarzschild and Kerr spacetimes} \label{s2a}

In order to demonstrate the reader that the notion of surface gravity could be obtained from the expression for the tidal acceleration, we provide an illustration for the Schwarzschild and Kerr spacetimes, which are two of the most important solutions of the Einstein field equations.

Let us first consider the Schwarzschild spacetime, which describes the gravitational field of a non-rotating spherically symmetric mass. For a comoving observer separated by a distance $\zeta^{\hat \rho}$ in the radial direction, the tidal acceleration is given by~\cite{MTW:73}
\begin{equation}
\frac{D^2\zeta^{\hat \rho}}{D\tau^2}= - R_{\hat \tau \hat \rho \hat \tau \hat \rho} \zeta^{\hat \rho}= \frac{2 M}{r^3} \zeta^{\hat \rho},
\end{equation}
where $M$ is the black hole mass, $r$ is the radial coordinate, and $\tau$ is the proper time along the trajectory of the particles. We can consider the case where the initial separation $\zeta^{\hat \rho}$ is infinitesimal in the radial direction $dr$, which gives the infinitesimal tidal acceleration in the radial direction $d\kappa$. We can then calculate the tidal acceleration by integrating this infinitesimal quantity from the horizon to infinity
\begin{equation}
\kappa= \int d\kappa= \int_{2 M}^\infty \frac{2 M}{r^3} \delta r= \frac{1}{4 M}.
\end{equation}
Thus, we have recovered the usual surface gravity of Schwarzschild black hole.

It is important to note that there are numerous trivial ways of obtaining $1/4 M$ in Schwarzschild spacetime. Therefore, one may not pursue this idea of extracting surface gravity seriously unless verifying the procedure for some nontrivial example. For this reason, we turn to the Kerr spacetime, which describes the gravitational field of a rotating black hole. Consider a comoving observer moving along the axis of rotation and separated by a distance $\zeta^{\hat \rho}$ in the radial direction. For this observer, the tidal acceleration is given by~\cite{jch9}
\begin{equation}
\frac{D^2\zeta^{\hat \rho}}{d\tau^2}= - R_{\hat \tau \hat \rho \hat \tau \hat \rho} \zeta^{\hat \rho}= \frac{2 M r}{(r^2+ a^2)^3} (r^2- 3 a^2) \zeta^{\hat \rho},
\end{equation}
where $a$ is the spin parameter of the black hole. Again, we can consider the case where the initial separation $\zeta^{\hat \rho}$ is infinitesimal in the radial direction $dr$, which gives the infinitesimal tidal acceleration in the radial direction $d\kappa$. The tidal acceleration results from the integration of this infinitesimal quantity from the horizon to infinity
\begin{align}
    \kappa= \int_{r_+}^\infty \frac{2 M r}{(r^2+ a^2)^3} (r^2- 3 a^2) \delta r= \frac{\sqrt{M^2- a^2}}{a^2+ r_+^2},
\end{align}
where $r_+= M+ \sqrt{M^2- a^2}$ is the radius of the event horizon. This gives the familiar surface gravity for Kerr black hole. In taking the example of Kerr metric, we have not gone outside our domain of spherical symmetry. This is due to the fact that the event horizon of the Kerr metric is itself spherical, and the calculations are restricted to the equatorial plane where the equations of motion are very similar to those for spherical symmetry. Principal trajectories of Kerr spacetime, in general, are also independent of $\theta$ coordinate~\cite{sc15}.

\subsection{Formal developments} \label{fd}

In this section, we develop a generic formalism to calculate the tidal surface gravity from the tidal acceleration between two freely falling particles in curved spacetimes. Consider a pair of such particles separated by a distance $\zeta^{\rho}$, the relative acceleration they experience can be expressed as \cite{wb3}
\begin{equation}
\frac{D^2\zeta^{\mu}}{D\tau^2}= - R^\mu_{\nu \lambda \rho} \zeta^{\lambda} \frac{dx^\nu}{d\tau}
\frac{dx^\rho}{d\tau}, \label{eq1}
\end{equation}
where $R^\mu_{\nu \lambda \rho}$ is the Riemann tensor.

To facilitate the calculation, we introduce a parallel propagated orthonormal tetrad $(l^\mu, n^\mu, u^\mu, v^\mu)$, such that $l^\mu= dx^\mu/d\tau$ is the tangent vector to the trajectory of one of the particles. In this frame, the tidal acceleration takes the form
\begin{equation}
k^{\mu}_{~~;\nu} l^\nu= - R^\mu_{~~\nu \lambda \rho} \zeta^{\lambda} l^\nu l^\rho, \label{eq2}
\end{equation}
where $k^\mu := D\zeta^\mu/D\tau$.

We can define an infinitesimal scalar $d\kappa_i= e^\mu_{~~i} k_{\mu;\nu} l^\nu$ corresponding to the infinitesimal four-separation $\zeta^{\rho}$, which allows us to write this equation as
\begin{equation}
d\kappa_i= R_{\mu \nu \lambda \rho} \zeta^{\lambda} l^\nu l^\rho e^\mu_{~~i}. \label{eq3}
\end{equation}
Writing this scalar in the four-dimensional form is trivial as we can see from this equation that $d\kappa_0=0$ identically.

Let us write an arbitrary infinitesimal separation $\zeta^\mu$ between two particles momentarily at rest in the local inertial frame of an infalling observer as
\begin{equation}
\zeta^{i}= \left(-A_2 \delta r, A_3 \delta\theta, A_4\delta\phi\right), \label{eq4}
\end{equation}
for some arbitrary functions $A_2$, $A_3$ and $A_4$. We note that $\zeta^{i}$ is tangent to the spacelike geodesic starting from an arbitrary point of $\gamma$ and orthogonal to it
\begin{equation}
    e^\mu_{~~i} \zeta^{i} l_\mu\Big |_{\gamma}=0,
\end{equation}
where $\gamma$ is the worldline of one of the particles. To obtain this equality, we have used the relation
\begin{multline}
    \zeta^\mu= A_1 l^\mu \delta t +e^\mu_{~~i} \zeta^{i}\\
    = A_1 l^\mu \delta t- A_2 n^\mu \delta r+ A_3 u^\mu \delta\theta+ A _4 v^\mu \delta\phi, \label{sv109}
\end{multline}
for some arbitrary function $A_1$. So, by definition, $\zeta^{i}$ constitute the Fermi normal coordinates \cite{mm20,n21}, while $\zeta^{\mu}$ still gives an arbitrary infinitesimal separation.

Let us utilize the observer dependence property of surface gravity, which enables us to select a foliation generated by the vectors $(u^\mu, v^\mu)$. The normal to this two-surface is given by the vector $(l^\mu, n^\mu)$. Thus, for $\zeta^\mu$ defined normal to this foliation, we could write $\zeta^\mu= B_1 n^\mu + B_2 l^\mu$, for arbitrary functions $B_1$ and $B_2$. By comparing this expression with Eq.~\eqref{sv109} gives $\zeta^\mu= A_1 l^\mu \delta t -A_2 n^\mu \delta r$. Substituting this form of $\zeta^\mu$ back into Eq.~\eqref{eq3} yields the infinitesimal quantity
\begin{equation}
    d\kappa_i= - A_2 R_{i \hat\tau \hat\rho \hat\tau} \delta r= - A_2 R_{\hat\tau i \hat\tau \hat\rho} \delta r.
\end{equation}
Integrating this equation results in the tidal surface gravity
\begin{equation}
\kappa_i|_{\infty} - \kappa_i|_{r} = -\int_{r}^\infty A_2 R_{\hat\tau i \hat\tau \hat\rho} \delta r. \label{tsg12}
\end{equation}
We thus obtain an invariant notion of surface gravity that depends on the curvature of the spacetime. Also note that
\begin{equation}
{\cal K}= \frac{R_{\mu \nu \lambda \rho} n^{\lambda} l^\nu l^\rho n^\mu}{(g_{\lambda\mu} g_{\rho\nu}- g_{\lambda\nu} g_{\rho\mu}) n^{\lambda} l^\nu l^\rho n^\mu}= -R_{\hat\tau \hat\rho \hat\tau \hat\rho},
\end{equation}
is the Gaussian curvature of the two-surface generated by the vectors $(l^\nu, n^\nu)$. The use of this quantity to define an invariant gravitational density originated from the work of Synge~\cite{js14}. Here, the Gaussian curvature arises from the fact that in defining the surface gravity, we take a radial/principal trajectory from infinity to the horizon. This trajectory defines a curvature on two-dimensions $(l^\nu, n^\nu)$.

Among the components $\kappa_i$, it is important to note that for the case of spherical symmetry, which we are restricting ourselves to here, only $\kappa_2$ is relevant. This is due to the fact that, among the components of the curvature scalar $R_{\hat\tau i \hat\tau \hat\rho}$, only $R_{\hat\tau \hat\rho \hat\tau \hat\rho}$ is nonvanishing for the case of spherical symmetry.

The definition of surface gravity presented above relies on the use of two orthogonal vectors, namely $(l^\mu, n^\mu)$. In a four-dimensional spacetime ${\cal V}$, we can include two additional orthogonal vectors, completing the tetrad as $(l^\mu, n^\mu, u^\mu, v^\mu)$. Petrov D spacetimes allow for a general procedure to solve the parallel transport equations~\cite{pd6,kf34}. Such spacetimes allow a unique $1+1+2$ decomposition, such that the first two components align with the principal vectors and the remaining two-surface is the surface orthogonal to the principal vectors~\cite{cb27}. Therefore, the definition of surface gravity given above yields a unique result in these spacetimes.

\subsubsection{Newtonian Limit}

In order to analyze the Newtonian limit of our definition of surface gravity, we first introduce the concept of Fermi normal coordinates. The geodesic curve $\gamma$ with tangent $l^\mu$ representing the motion of a test particle together with the tetrad $e^\mu_{~~i}= (n^\mu, u^\mu, v^\mu)$ form a comoving local reference frame. The position of a point of a star with respect to the chosen reference frame can be described by a three-vector $\vec{\zeta}= \sum_{i=1}^3 \zeta^i e_i$, and in the Fermi normal coordinates $T, \zeta^i$, with $T=\hat \tau$, the line element up to quadratic order can be written in the form~\cite{km19}
\begin{multline}
ds^2= -\left(1+ R_{\hat\tau i \hat\tau j} \zeta^i \zeta^j\right) dT^2- \frac{4}{3} R_{\hat\tau i j k} \zeta^i \zeta^k dT d\zeta^j\\
+ \left(\delta_{i j}- \frac{1}{3} R_{i k j l} \zeta^k \zeta^l\right) d\zeta^i d\zeta^j.
\end{multline}
In the Newtonian limit, where the gravitational field could be approximated from the weak static source, the metric component $g_{00}$ can be written in terms of the Newton potential as $g_{00}= -(1+ 2\Phi)$, and we obtain the component of the curvature tensor $R_{\hat\tau i \hat\tau j}= \Phi_{,i j}$. In this limit, the geodesic deviation equation takes the form~\cite{fn17}
\begin{equation}
\int\frac{d^2 \zeta_i}{d\tau^2}= -\int R_{\hat\tau i \hat\tau j} \zeta^j= -\Phi_{,i}.
\end{equation}
We thus have the tidal surface gravity $\kappa_i= -\int R_{\hat\tau i \hat\tau j} \zeta^j= - \Phi_{,i}$, which coincides with the acceleration due to gravity in the Newtonian limit.

\subsection{General Structure of the Deviation Vector} \label{s3}

The deviation vector given in Eq.~\eqref{sv109}, in a particular choice of spacetime foliation, contains two arbitrary and undetermined functions. In this section, we will adopt the approach presented in Ref.~\cite{B:16} to discuss the general structure of the deviation vector, in order to constrain these undetermined functions.

We consider a family of geodesics, given by $x^\alpha= x^\alpha(\tau, s)$, where the parameter $s$ labels the individual geodesic, parameterized by proper time $\tau$. The tangent vectors to these curves are defined as $l^\alpha= \partial x^\alpha/\partial\tau$ and the deviation vector as $\zeta^{\alpha}= \delta x^\alpha=\partial x^\alpha/\partial s$. The covariant derivatives along the curves are denoted by $D_\tau$ and $D_s$, respectively.

The evolution and deformation of the deviation vector along the geodesic is given by the following equation:
\begin{equation}
D_\tau \zeta^\alpha= \zeta^\beta \nabla_{\beta} l^\alpha. \label{dd16}
\end{equation}
This equation follows from the fact that second partial derivatives with respect to $\tau$ and $s$ commute.

As a consequence of this equation and of the fact that $l^\alpha \nabla_\beta l_\alpha=0$, we have
\begin{equation}
    l_\alpha D_\tau \zeta^\alpha= D_\tau l_\alpha \zeta^\alpha=0 \implies l_\alpha \zeta^\alpha= \mathrm{Const.}
\end{equation}
In the following calculations, we will use these two properties of the deviation vector to determine the surface gravity explicitly.

\subsubsection{Example in (t,r) coordinates}

In this section, we use the approach outlined in the previous section to calculate the deviation vector in (t,r) coordinates.

Assuming that the deviation vector has the form given in Eq.~\eqref{sv109}, simplified for the particular choice of spacetime foliation, i.e.,
\begin{equation}
\zeta^\mu= \left(-A_2 \dot r n^\mu+ A_1 \dot t l^\mu \right) d\tau, \label{dv18}
\end{equation}
we first impose the condition $\zeta^\alpha l_\alpha= \mathrm{Const.}$, which gives the relation
\begin{equation}
A_1 \dot t= \mathrm{Const.}= {\cal D}.
\end{equation}

Next, we calculate $l^\mu \nabla_\mu \zeta^\alpha$ and $\zeta^\mu \nabla_\mu l^\alpha$ and use the equation for the evolution and deformation of the deviation vector (Eq.~\eqref{dd16}) to obtain
\begin{equation}
n^\alpha l^\mu \nabla_\mu \left(A_2 \dot r \right)= A_2 \dot r n^\mu \nabla_\mu l^\alpha.
\end{equation}
Simplifying this expression yields the value for $A_2$, which is required for the calculation of the surface gravity:
\begin{equation}
A_2= \frac{1}{\dot r} \exp\left( \int n^\alpha n^\mu \nabla_\mu l_\alpha d\tau\right). \label{a21}
\end{equation}

Thus, we have demonstrated a generic procedure for determining the deviation vector in non-stationary spacetimes and used this procedure to obtain a specific expression for the deviation vector in (t,r) coordinates. This form of deviation vector is necessary for calculating the surface gravity in non-stationary spacetimes.

\subsection{Application to stationary spacetimes} \label{s4}

\subsubsection{General setup}

Before delving into the specifics of stationary spacetimes, it is useful to establish a calculation that holds for general spherical symmetry. To begin our calculation, we define principal vectors $l^\mu$ and $n^\mu$ as
\begin{align}
    l^\mu=& (\dot t, \dot r, 0, 0),\label{l113}\\
    n^\mu=& \left(\frac{1}{e^h f} \dot r, e^h f \dot t, 0, 0\right), \label{n114}
\end{align}
where
\begin{equation}
    - e^{2 h(t,r)} f(t,r) \dot t^2+ \frac{1}{f(t,r)} \dot r^2= -1.
\end{equation}
Then, we solve for the curvature component $R_{\hat\rho \hat\tau \hat\rho \hat\tau}$ in its frame. This yields the expression
\begin{multline}
    R_{\hat\rho \hat\tau \hat\rho \hat\tau}= R_{\mu\nu\alpha\beta} l^\mu n^\nu l^\alpha n^\beta\\
= \frac{e^{-h}}{2}\frac{\partial}{\partial r} \left(e^{-h} \frac{\partial (e^{2 h} f)}{\partial r}\right)+ \frac{e^{-h}}{2} \frac{\partial}{\partial t} \left(e^{- h} f^{-2} \frac{ \partial f}{\partial t}\right), \label{ct25}
\end{multline}
which we can further simplify using the assumption that $f$ and $e^h$ depend on time only through the Misner-Sharp mass $M(t,r)$~\cite{ms13,ms14}. We work in the semiclassical limit, where we assume small evaporation rate and thus neglect terms of the form $(\partial M/\partial t)^2$ and $\partial^2M/\partial t^2$. We then integrate to obtain
\begin{equation}
\kappa= - \int_{r_H}^\infty A_2 \frac{e^{-h}}{2}\frac{\partial}{\partial r} \left(e^{-h} \frac{\partial (e^{2 h} f)}{\partial r}\right) \delta r, \label{tsg26}
\end{equation}
where $r_H$ is the event/apparent horizon radius and the general form of $A_2$ is given in Eq.~\eqref{a21}. This expression gives the tidal surface gravity for general spherically symmetric spacetimes in Schwarzschild coordinates, where a particular choice of foliation is adopted. It is worth noting, however, that a different choice of foliation would be preferred when calculating surface gravity in advanced/retarded Eddington-Finkelstein coordinates (appendix~\ref{appa} provides a detailed calculation of the surface gravity in advanced coordinates).

\subsubsection{Stationary spacetimes}

Now, we apply the previous results to stationary spacetimes. Specifically, we use Eq.~\eqref{tsg26}, which follows identically from Eq.~\eqref{ct25} without any approximation. This is because, in stationary spacetimes, the time derivative terms present in the final equality of Eq.~\eqref{ct25} identically vanishes. To determine $A_2$, we consider the deviation vector $\zeta^\mu$ defined in Eq.~\eqref{dv18}
\begin{equation}
\zeta^\mu= \left(A_1 \dot t^2- A_2 \frac{\dot r^2}{e^h f}, A_1 \dot t \dot r- A_2 e^h f \dot t \dot r, 0, 0 \right) d\tau.
\end{equation}
where Eqs.~\eqref{l113} and \eqref{n114} are used for simplification. Demanding that $\zeta^\mu$ be a timelike Killing vector gives $A_1 = A_2 e^h f$. As we are using a Killing vector as a deviation vector, it is necessary to show that the Killing vector also satisfies the tidal force equation, which we do in appendix~\ref{appb}. Substituting this relation back into the equation for $\zeta^\mu$ and setting $A_2 = e^h$, we get
\begin{equation}
\zeta^\mu = \left(1, 0, 0, 0 \right) d\tau,
\end{equation}
which satisfies the timelike Killing vector condition. We use this expression for $A_2$ into Eq.~\eqref{tsg26} to obtain
\begin{equation}
\kappa= -\frac{e^{-h}}{2} \frac{\partial (e^{2 h} f)}{\partial r}\bigg|_{r=r_H}, \label{sg38}
\end{equation}
where, we have assumed asymptotic flatness at infinity. Further simplification gives
\begin{equation}
    \kappa= -\frac{e^h}{2} \left(2 f \frac{\partial h}{\partial r}+ \frac{\partial f}{\partial r}\right)\bigg|_{r=r_H}= -\frac{e^h}{2} \frac{\partial f}{\partial r}\bigg|_{r=r_H},
\end{equation}
which exactly matches the result for the (Peeling) surface gravity given in Ref.~\cite{av15}. In obtaining the last equality, we have applied the same approximation used in the given reference. Specifically, we have taken $\partial h/\partial r$ to be finite, such that the expression containing it can be neglected with respect to the remaining terms in the equation (this assumption is not made anywhere else).

\subsubsection{More on Thermodynamics}

The study of black hole thermodynamics is of great importance for understanding the nature of gravity and its relationship with thermodynamics. In this section, we will use the generalized expression for surface gravity derived in the previous section to study the laws of thermodynamics for black holes.

The surface gravity of a black hole plays a crucial role in black hole thermodynamics, as it is directly related to the Hawking temperature, which governs the thermal properties of the black hole. In the previous section, we derived a generalized expression for the surface gravity, which can be written alternatively in the form of an antisymmetric second rank tensor $F_{\rho\mu}$ as (see appendix~\ref{appb} for the derivation)
\begin{equation}
    \kappa= F_{\rho\mu} l^\rho n^\mu,
\end{equation}
where
\begin{equation}
    F_{\rho\mu}= \frac{1}{2}\left(\zeta_{\mu;\rho}- \zeta_{\rho;\mu}\right).
\end{equation}
Using this expression, we can write the first law of black hole thermodynamics as
\begin{equation}
    M= \frac{1}{4\pi} \lim_{S_t \to \infty}\oint_{S_t} F_{\rho\mu} l^\rho n^\mu \sqrt{\sigma} d^2\theta,
\end{equation}
where $\sigma_{AB}$, $A,B= 1,2$ is the metric on the two-surface $S_t$. The first law gives a relation for the change in energy of a black hole in terms of its change in area and surface gravity. We thus have the Komar formula relating the mass of the black hole to the flux of the Komar current, which is conserved along the horizon~\cite{ep16}.

If we consider that the boundary $S_t$ of the hypersurface $\Sigma$ consists of $\partial \Sigma$, we can rewrite this relation for mass as~\cite{wb13,bch16}:
\begin{equation}
M = \frac{1}{4\pi} \int_{\Sigma} F^{\rho\mu}_{~~~~;\mu} l_\rho \sqrt{\gamma^{(\Sigma)}} d^3x  + \frac{1}{4\pi} \oint_{\partial \Sigma} F_{\rho\mu} l^\rho n^\mu \sqrt{\sigma} d^2\theta,
\end{equation}
where $d\Sigma_\mu= l_\rho \sqrt{\gamma^{(\Sigma)}} d^3x$ is the surface element of $\Sigma$. The first term on the right-hand side of this equation can be simplified using the equation for the Killing vector:
\begin{equation}
    F^{\rho\mu}_{~~~~;\mu} l_\rho= -R_{\beta\alpha} l^\beta \xi^\alpha.
\end{equation}
Using this equation, we get
\begin{equation}
M = -\frac{1}{4\pi} \int_{\Sigma} R_{\beta\alpha} l^\beta \xi^\alpha \sqrt{\gamma^{(\Sigma)}} d^3x + \frac{1}{4\pi} \oint_{\partial \Sigma} F_{\rho\mu} l^\rho n^\mu \sqrt{\sigma} d^2\theta,
\end{equation}
which gives the first law for the stationary spacetimes.

\subsection{Spacetimes conformal to stationary spacetimes} \label{s5}

To compute the tidal forces in these spacetimes, we will first show that in the limit of our approximation where we neglect terms of the form $(\partial M/\partial t)^2$ and $\partial^2M/\partial t^2$, conformal Killing vectors also satisfy the tidal force equation. For this, we use the commutation relation for the covariant derivative
\begin{equation}
    \zeta_{\sigma;\rho;\mu}- \zeta_{\sigma;\mu;\rho}= - R^\lambda_{\sigma\rho\mu} \zeta_\lambda. \label{cr70}
\end{equation}
Exploiting the cyclic property of the curvature tensor
\begin{equation}
    \zeta_{\sigma;\rho;\mu}- \zeta_{\sigma;\mu;\rho}+ \zeta_{\mu;\sigma;\rho}- \zeta_{\mu;\rho;\sigma}+ \zeta_{\rho;\mu;\sigma}- \zeta_{\rho;\sigma;\mu}= 0,
\end{equation}
and the definition of the conformal Killing vector
\begin{equation}
    \zeta_{\nu;\mu}+ \zeta_{\mu;\nu}= {\cal B} g_{\mu\nu},
\end{equation}
for some arbitrary function ${\cal B}$, we obtain the following equation
\begin{equation}
    2 \zeta_{\mu;\rho;\sigma}= -{\cal B}_{;\mu} g_{\sigma\rho} + {\cal B}_{;\rho} g_{\mu\sigma} + {\cal B}_{;\sigma} g_{\rho\mu} - 2 R^\lambda_{\sigma\rho\mu} \zeta_\lambda.
\end{equation}
From this equation, one can infer that conformal Killing vectors do not satisfy tidal force equation, in general
\begin{equation}
    2 \zeta_{\mu;\rho;\sigma} e^\mu_{~~i} l^\rho l^\sigma= - 2 R^\lambda_{\sigma\rho\mu} \zeta_\lambda e^\mu_{~~i} l^\rho l^\sigma +{\cal B}_{;\mu} e^\mu_{~~i}.
\end{equation}
However, $\zeta^\mu- {\cal D}^\mu$, where $2 {\cal D}_{\mu;\rho;\sigma}= {\cal B}_{;\mu} g_{\rho\sigma}$ satisfies the tidal force equation
\begin{equation}
    \left(\zeta_\mu- {\cal D}_{\mu}\right)_{;\rho;\sigma}e^\mu_{~~i} l^\rho l^\sigma=- R^\lambda_{\sigma\rho\mu} \zeta_\lambda e^\mu_{~~i} l^\rho l^\sigma.
\end{equation}
Hence, the surface gravity for spacetimes conformal to stationary spacetimes, in general, could be determined from the fact that in such spacetimes, $\zeta_\mu- {\cal D}_\mu$ is the deviation vector. However, as ${\cal B}$ vanishes for stationary spacetimes, for small evaporation/accretion rates, the function ${\cal B}= \zeta^\mu_{~~;\mu}/2$, introduced above, could be expanded in terms of $\partial M/\partial t$. We thus have, for appropriate normalization of the conformal Killing vector
\begin{equation}
    \frac{\partial {\cal B}}{\partial x^\mu}= {\cal O}\left(\frac{\partial M}{\partial t}\right)^2+ {\cal O}\left(\frac{\partial^2 M}{\partial t^2}\right).
\end{equation}
So, neglecting terms containing $(\partial M/\partial t)^2$ and $\partial^2M/\partial t^2$, ${\cal B}$ would be constant, and the conformal Killing vector would satisfy the tidal force equation. These generic lines of reasoning will be illustrated through a specific example of advanced Vaidya spacetime, whose near horizon provides an excellent limit to the general spherically symmetric evaporating black holes~\cite{pf1}.

\subsubsection{Specific example of Vaidya}

The Vaidya spacetime has been extensively studied in the context of gravitational collapse and black hole physics. In this paper, we consider the Vaidya spacetime in advanced Eddington-Finkelstein coordinates
\begin{equation}
ds^2= - f(v,r) dv^2+ 2 dv dr+ r^2 d\Omega^2.
\end{equation}
We focus on the linear Vaidya case, where the metric is conformal to a static metric with a Killing vector that gives the temperature in the original spacetime. Accounting for the conformal factor gives the temperature of Vaidya~\cite{tj3}.

We will now calculate the surface gravity of linear Vaidya using conformal Killing vectors. The linear Vaidya metric could be written alternatively as conformal to the static spacetime
\begin{multline}
    ds^2= e^{-2\alpha T/r_0} \bigg(-\left(1-\frac{r_0}{R}+ \frac{2\alpha R}{r_0}\right) dT^2+ 2 dT dR\\
    + R^2 d\Omega^2\bigg),
\end{multline}
where $\alpha$ and $r_0$ are constants. The metric is conformal to the static spacetime. The transformation relations between the advanced $(v,r)$ and the static $(T,R)$ coordinates are given by
\begin{align}
\alpha v=& r_0 - r_0 e^{-\alpha T/r_0}, \qquad dv= e^{-\alpha T/r_0} dT \nonumber\\
r=& R e^{-\alpha T/r_0}, \qquad dr= e^{-\alpha T/r_0}\left(dR- \frac{\alpha R}{r_0} dT\right).
\end{align}
The conformal Killing vector of the Vaidya metric is given by
\begin{equation}
    \zeta^\mu= -\left(\frac{M}{M'}, r,0,0\right).
\end{equation}
In static $(T,R)$ coordinates, it becomes
\begin{equation}
    \zeta^\mu= \left(\frac{r_0}{\alpha}, 0,0,0\right).
\end{equation}
From this, we see that the conformal factor ${\cal B}= \mathrm{Const.}$, from which we could write ${\cal D}=0$. To match the deviation vector given in Eq.~\eqref{dv39} with the conformal Killing vector, we should have the following values of $A_1$ and $A_2$
\begin{equation}
A_1= \left(A_2- \frac{r}{f \dot v \dot r- \dot r^2}\right) f, \quad A_2= \frac{r \dot v}{\dot r}- \frac{M}{M'}.
\end{equation}
Also in $(T,R)$ coordinates, we could write
\begin{equation}
    A_2 \dot r= \frac{r_0}{\alpha} e^{-2 \alpha T/r_0} \dot R.
\end{equation}

We can calculate the tidal acceleration between two freely moving particles momentarily at rest in the local inertial frame of an infalling observer using the formula
\begin{equation}
\frac{D^2\zeta^{\hat \rho}}{d\tau^2}= - R_{\hat \tau \hat \rho \hat \tau \hat \rho} \zeta^{\hat \rho}= \frac{2 M}{r^3} \zeta^{\hat \rho}.
\end{equation}
It should also be emphasized here that the component of the curvature tensor in the local inertial frame freely falling with the observer $R_{\hat \tau \hat \rho \hat \tau \hat \rho}$ is equal to the component of the curvature tensor in the locally inertial static orthonormal frame $R_{\hat t \hat \rho \hat t \hat \rho}$. Taking $\zeta^{\hat \rho}= - A_2 dr$ from Eq.~\eqref{sv109}, we obtain the surface gravity
\begin{equation}
    \kappa= \int d\kappa= \frac{\Omega \alpha}{r_0} \int_{2 M}^\infty A_2 \frac{2 M}{r^3} \delta r = \frac{1}{2 M} \int \frac{r_0^2}{R^3} dR= \frac{1}{4 M},
\end{equation}
where $\Omega^2= e^{2\alpha T/r_0}$ is the conformal factor, which transforms the Vaidya metric into a static spacetime. We have rescaled the surface gravity, as this is necessary to make the physical prediction equivalent in two conformal frames~\cite{fl36}. Moreover, the prefactor $\alpha/r_0$ is taking care of the normalization of the Killing vector in static $(T,R)$ coordinates. The term containing $M'$ is absent in the surface gravity of Vaidya. Note that the term $R_{\hat \tau \hat \rho \hat \tau \hat \rho}$ is coordinate independent and the radial coordinate $r$ in Schwarzschild and Vaidya are the same. Hence, although this result is convenient to obtain in $(v,r)$ coordinates, it would be unchanged if we use Schwarzschild coordinates.

Moreover, despite being a coordinate-independent quantity, surface gravity is actually observer-dependent. While working, for example, in advanced $(v,r)$ coordinates, the foliation generated by the $v=\mathrm{const.}$ surface is usually preferred. However, it is worth mentioning that this choice of foliation generally does not coincide with the foliation generated by the $t=\mathrm{const.}$ surface in Schwarzschild $(t,r)$ coordinates.

\section{Discussion and conclusions} \label{s6}

We have reviewed various notions of dynamical surface gravity and presented a method for calculating the surface gravity from tidal acceleration. The calculation of tidal surface gravity depends on the spacetime curvature, would-be-horizon and an observer. These are all the features surface gravity should have if we want to relate it to the Hawking temperature. We have seen that this procedure gives the correct surface gravity for the standard cases of Schwarzschild and Kerr spacetimes. Moreover, we were able to retrieve the (Peeling) surface gravity from the literature in the general case of stationary spacetimes in spherical symmetry.

The term containing $M'(v)$ does not appear in the surface gravity of Vaidya shows that small dynamical effects do not contribute significantly to the Hawking process~\cite{pf1}. To obtain this result we have integrated from the conformal Killing horizon, which was shown to coincide with the Rindler horizon. This integration from the conformal Killing horizon is reflected in the normalization factor, because of which the time component of the conformal Killing vector reduces to unity at the horizon (only the value at horizon contributes to the surface gravity).

It is interesting to see how this notion of surface gravity generalizes to the axially symmetric spacetimes. Only Petrov D spacetimes admit the additional constant of motion, which allow complete integrability of geodesic equations and analytic solution of the equation of parallel transport equation applied to an orthonormal tetrad~\cite{kf34,pd6}. These are the necessary quantities for calculating tidal acceleration and hence surface gravity.

\section*{Acknowledgements}

PKD is supported by an International Macquarie University Research Excellence Scholarship.





\appendix

\section{Stationary spacetimes in (v,r) coordinates \label{appa}}

Analogous to Eqs.~\eqref{l113} and \eqref{n114}, we define principal null vectors in $(v,r)$ coordinates as
\begin{align}
    l^\mu=& (\dot v, \dot r, 0, 0),\label{vl113}\\
    n^\mu=& \left(\dot v, - \dot r+ e^{h_+} f \dot v, 0, 0\right), \label{vn114}
\end{align}
where
\begin{equation}
    - e^{2 h_+(v,r)} f(v,r) \dot v^2+ 2 e^{h_+(v,r)}\dot v \dot r= -1.
\end{equation}
Then, with this choice of tetrad, we write the curvature component $R_{\hat\rho \hat\tau \hat\rho \hat\tau}$ in its frame
\begin{multline}
    R_{\hat\rho \hat\tau \hat\rho \hat\tau}= R_{\mu\nu\alpha\beta} l^\mu n^\nu l^\alpha n^\beta\\
    = e^{-h_+} \left(\frac{1}{2} \frac{\partial }{\partial r}\left(e^{h_+} \frac{\partial f}{\partial r}\right)+ \frac{\partial }{\partial r}\left(e^{h_+} \frac{\partial h}{\partial r} f \right)+ \frac{\partial^2 h_+}{\partial v \partial r}\right). \label{ct59}
\end{multline}
We thus can substitute this relation into Eq.~\eqref{tsg12} to obtain the tidal surface gravity in $(v,r)$ coordinates. Now, in stationary spacetimes, the time derivative terms present in the final equality of Eq.~\eqref{ct59} identically vanishes. To determine $A_2$, we consider the deviation vector $\zeta^\mu$ defined in Eq.~\eqref{dv18} in $(v,r)$ coordinates
\begin{align}
    \zeta^\mu=
    & \Bigg(A_1 e^{h_+} \dot v^2- \frac{A_1}{f} \dot v \dot r- A_2 \dot v \dot r, \nonumber\\
    & A_1 e^{h_+} \dot v \dot r- \frac{A_1}{f} \dot r^2+ A_2 \dot r^2- A_2 e^{h_+} f \dot v \dot r, 0, 0 \Bigg) d\tau, \label{dv39}
\end{align}
where Eqs.~\eqref{vl113} and \eqref{vn114} are used for simplification. The factor $e^{h(t,r)}$, which appears from coordinate transformation, is absorbed into the coefficient $A_1$, which is not important for later calculations. Demanding that $\zeta^\mu$ be a timelike Killing vector gives $A_1= A_2 f$. Substituting this relation back into the equation for $\zeta^\mu$ and setting $A_2= e^{h_+}$, we get
\begin{equation}
    \zeta^\mu= \left(1, 0, 0, 0 \right) d\tau,
\end{equation}
which satisfies the timelike Killing vector condition. We use this expression for $A_2$ to obtain the following relation for the tidal surface gravity in $(v,r)$ coordinates
\begin{equation}
    \kappa= -\frac{e^{-h_+}}{2} \frac{\partial (e^{2 h_+} f)}{\partial r}\bigg|_{r=r_+},
\end{equation}
where we have assumed asymptotic flatness at infinity.

\section{Tidal force and Killing vector} \label{appb}

The Killing condition, $\zeta_{\mu;\nu}+ \zeta_{\nu;\mu}=0$, can be combined with the commutation rule for the covariant derivatives and the cyclic sum rule for the curvature tensor to show that a Killing vector satisfies the tidal force equation~\cite{wb3}
\begin{equation}
    \zeta_{\mu;\rho;\sigma}= - R^\lambda_{~~\sigma\rho\mu} \zeta_\lambda \implies \zeta_{\mu;\rho;\sigma} l^\rho l^\sigma= -R^\lambda_{~~\sigma\rho\mu} \zeta_\lambda l^\rho l^\sigma.
\end{equation}
So, the surface gravity, $d\kappa_i= e^\mu_{~~i} k_{\mu;\nu} l^\nu$, takes the form
\begin{equation}
    d\kappa_i= \frac{1}{2} \zeta_{\mu;\rho;\nu} \left(l^\rho e^\mu_{~~i}- e^\rho_{~~i} l^\mu \right) l^\nu= R_{\mu\nu\alpha\beta} l^\nu l^\beta e^\alpha_{~~i} \zeta^\mu.\label{sg33}
\end{equation}
In obtaining the second equality, we have used the Killing condition stated above. As the vectors $l^\rho$ and $e^\mu_{~~i}$ are parallel propagated, integrating this equation along the trajectory of $l^\mu$ yields the expression for the surface gravity:
\begin{equation}
    \kappa_i= \frac{1}{2} \zeta_{\mu;\rho} \left(l^\rho e^\mu_{~~i}- e^\rho_{~~i} l^\mu \right), \label{sg20}
\end{equation}
which is exactly the same as the expression given in Ref.~\cite{wb13}.

\end{document}